\documentclass[manuscript]{aastex}          
\usepackage{spr-astr-addons}    

\begin{document}
%
\title{Characteristic Count Rate Profiles for a\\Rotating Modulator Gamma-Ray Imager}

\shorttitle{Characteristic Profiles for an RM}
\shortauthors{Budden, Budden, Case, Cherry}

\author{Brent S. Budden}
\email{bbudden@phys.lsu.edu}
\affil{Dept. of Physics \& Astronomy, Louisiana State University, Baton Rouge, LA 70803}
\and
\author{Mark R. Budden} 
\affil{Dept. of Mathematics and Computer Science, Western Carolina University, Cullowhee, NC 28723}
\and
\author{Gary L. Case}
\and
\author{Michael L. Cherry} 
\affil{Dept. of Physics \& Astronomy, Louisiana State University, Baton Rouge, LA 70803}
\and 

\begin{abstract}
Rotating modulation is a technique for indirect imaging in the hard x-ray and soft gamma-ray energy bands, which may offer an advantage over coded aperture imaging at high energies. A rotating modulator (RM) consists of a single mask of co-planar parallel slats typically spaced equidistance apart, suspended above an array of circular non-imaging detectors. The mask rotates, temporally modulating the transmitted image of the object scene. The measured count rate profiles of each detector are folded modulo the mask rotational period, and the object scene is reconstructed using pre-determined characteristic modulation profiles. The use of Monte Carlo simulation to derive the characteristic count rate profiles is accurate but computationally expensive; an analytic approach is preferred for its speed of computation. We present both the standard and a new advanced characteristic formula describing the modulation pattern of the RM; the latter is a more robust description of the instrument response developed as part of the design of a wide-field high-resolution telescope for gamma-ray astronomy. We examine an approximation to the advanced formula to simplify reconstruction software and increase computational speed, and comment on both the inherent limitations and usefulness of the approach. Finally, we show comparisons to the standard formula and demonstrate image reconstructions from Monte Carlo simulations.
\end{abstract}

\keywords{Image Reconstruction, Gamma ray, X-ray}

\section{Introduction}

\begin{figure}[!b]
	\begin{center}
		\includegraphics[height=2.3in]{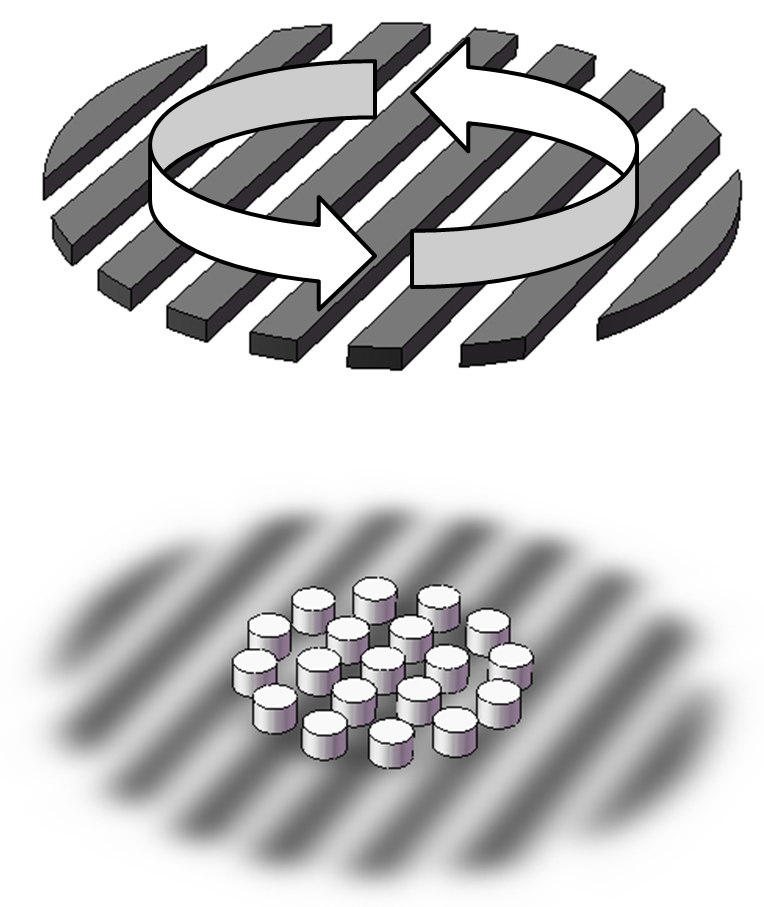}
	\end{center}
	\caption[example]
   { \label{fig:rm} 
   A rotating modulator consists of a single mask of opaque slats that rotates above an array of circular non-imaging detectors.}
\end{figure}

X-ray/gamma-ray imaging can be performed by temporal modulation techniques, whereby incident photons are encoded by a moving component of the instrument. By time-tagging each detected photon, knowledge of the instrument response can be used to reconstruct the object scene. A rotating modulation collimator (RMC) is one of the most common instruments of this class; it uses a single detector to measure the time variation of the counting rates produced by the interfering shadows of two rotating masks of finely-spaced opaque slats \citep{Mertz1967}. Despite the low mask transmission (25\%), and thus low sensitivity, RMCs have been used in rocket \citep{Schnopper1970}, balloon \citep{crannell1986, gaither1996}, and spacecraft \citep{hurford2002} observations at x-ray energies.

Sensitivity may be increased by utilizing a single mask design, which increases the transmission to 50\%. A rotating modulator (RM) \citep{Durouchoux1983, Dadurkevicius1985} is one such instrument developed to image hard x-ray and gamma-ray photons (tens of keV to MeV). As we have shown previously \citep{BuddenTNS2010}, the RM may have some significant sensitivity advantages over the commonly-used coded aperture, particularly at high energies.

The RM consists of a mask of co-planar parallel slats typically spaced equidistance apart rotating above an array of circular detectors (Fig. \ref{fig:rm}). The transmission of photons from the object scene, $S$, is modulated in time, and so a history of counts is recorded by each detector. For a stationary instrument, the recorded data are folded modulo the mask period to produce a count profile for detector $d$, $O_d(t)$, which may be described by
\begin{equation}
	O_d(t) = \sum_n P_d(t,n) S(n),
	\label{eq:oMatrix}
\end{equation}
if noise is ignored. $P_d(t,n)$ is the instrument response function, which in this application is a collection of characteristic count rate profiles for point sources at all possible scene locations $n$. Image reconstruction is the technique by which the inverse problem of Eq. \ref{eq:oMatrix} is solved for the object scene $S$. Obviously, the modulation patterns $P_d$ must be pre-determined and well-defined.

The ideal technique for determining the expected modulation patterns should be computationally fast, allow for unconstrained instrument geometry, account for projection effects and non-uniform attenuation, and describe the cumulative shadowing by multiple slats simultaneously. Brute force Monte Carlo simulations are able to accomplish these tasks, but the computation is time-consuming. \cite{Durouchoux1983} and \cite{Dadurkevicius1985} have presented a standard characteristic profile that can be calculated analytically, as described in Sec. \ref{sec:classical}. While suitable in many scenarios, this formula imposes tight constraints on instrument geometry and is too simplistic to account for non-uniform attenuation or shadow lengthening.

In this paper, the previous standard formula is extended and made more general, with particular care taken to account for incomplete mask absorption which becomes important at hard x-ray and gamma-ray energies. This analysis is part of a program to design a wide-field high resolution telescope suitable for hard x-ray/gamma-ray imaging from a long-duration balloon or satellite platform \citep{Grindlay2001, McConnell2004, BuddenSORMA2010}. To achieve a wide field of view (FOV) and sensitivity to high energies (which requires thick mask slats), a more robust analytical profile is necessary to accurately describe the instrument response.

In Sec. \ref{sec:advanced}, we present an advanced characteristic profile for the RM that is capable of describing this complex modulation pattern and can be calculated analytically in a relatively short time. In Sec. \ref{sec:results}, we show examples of count rate profiles generated with the standard and advanced formulae and with Monte Carlo simulations, as well as reconstructed images. The image reconstruction technique has been described elsewhere \citep{BuddenTNS2010}, and has been shown to be capable of providing coded-aperture quality resolution with better detector efficiency at high energies, resolving multiple closely-separated sources and operating in the presence of background.

\section{Standard Characteristic Count Rate Formula}
\label{sec:classical}

\begin{figure}[!t]
	\begin{center}
		\includegraphics[width=2.2in]{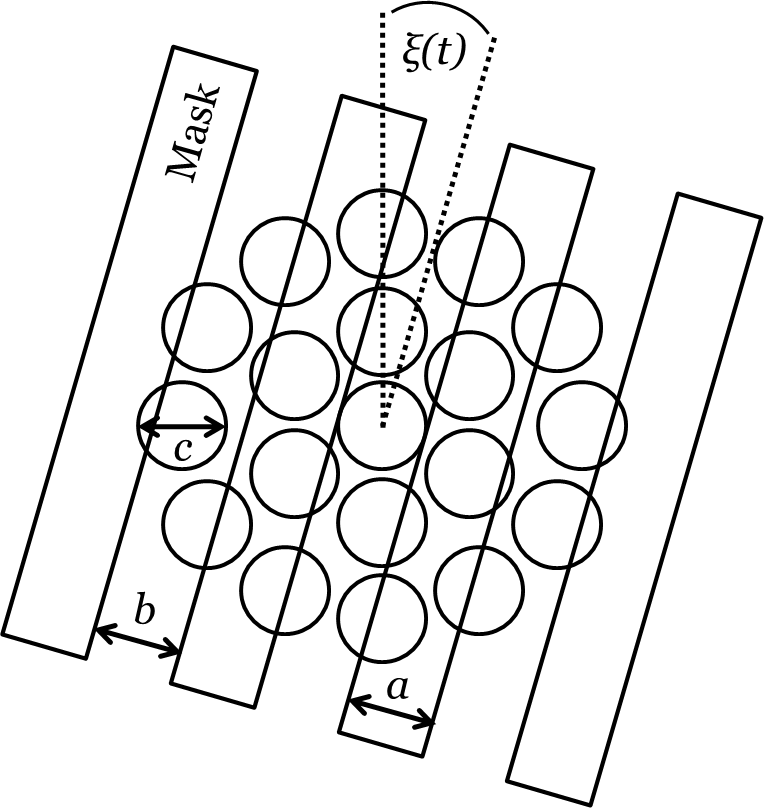} \\
	\end{center}
	\caption[example]
   { \label{fig:geometry} 
   Top view diagram of the RM geometry. A mask with slat width $a$ and slat spacing $b$ rotates above an array of circular detectors with diameter $c$, according to the function $\xi(t)$. For the standard formula (Sec. \ref{sec:classical}), $a = b = c$, but the advanced formula (Sec. \ref{sec:advanced}) allows these terms to be defined independently.}
\end{figure}

The standard characteristic formula for a single-mask RM was first presented by \cite{Durouchoux1983} and examined in greater detail by \cite{Dadurkevicius1985}. An RM has slat width $a$, slat spacing $b$, and detector diameter $c$ (Fig. \ref{fig:geometry}), and the mask is suspended a distance $L$ from the detection plane. The standard formula imposes the constraint $a = b = c$, maximizing instrument sensitivity and count rate profile contrast. The assumption is made that slats have infinitesimal thickness, but attenuate 100\% of the incident photons. An attenuation coefficient may be applied to correct for transmission through the slats, but clipping effects, non-uniform attenuation, and shadow lengthening are ignored. For the description of both the standard formula below and the advanced formula in Sec. \ref{sec:advanced}, the mask is assumed to be centered midway between two slats and begins its period with the bars parallel to the lab frame's $\hat{y}$ direction; a simple transformation and offset parameter, however, can easily provide an alternate case.

\begin{figure}[!t]
	\begin{center}
		\includegraphics[width=2.9in]{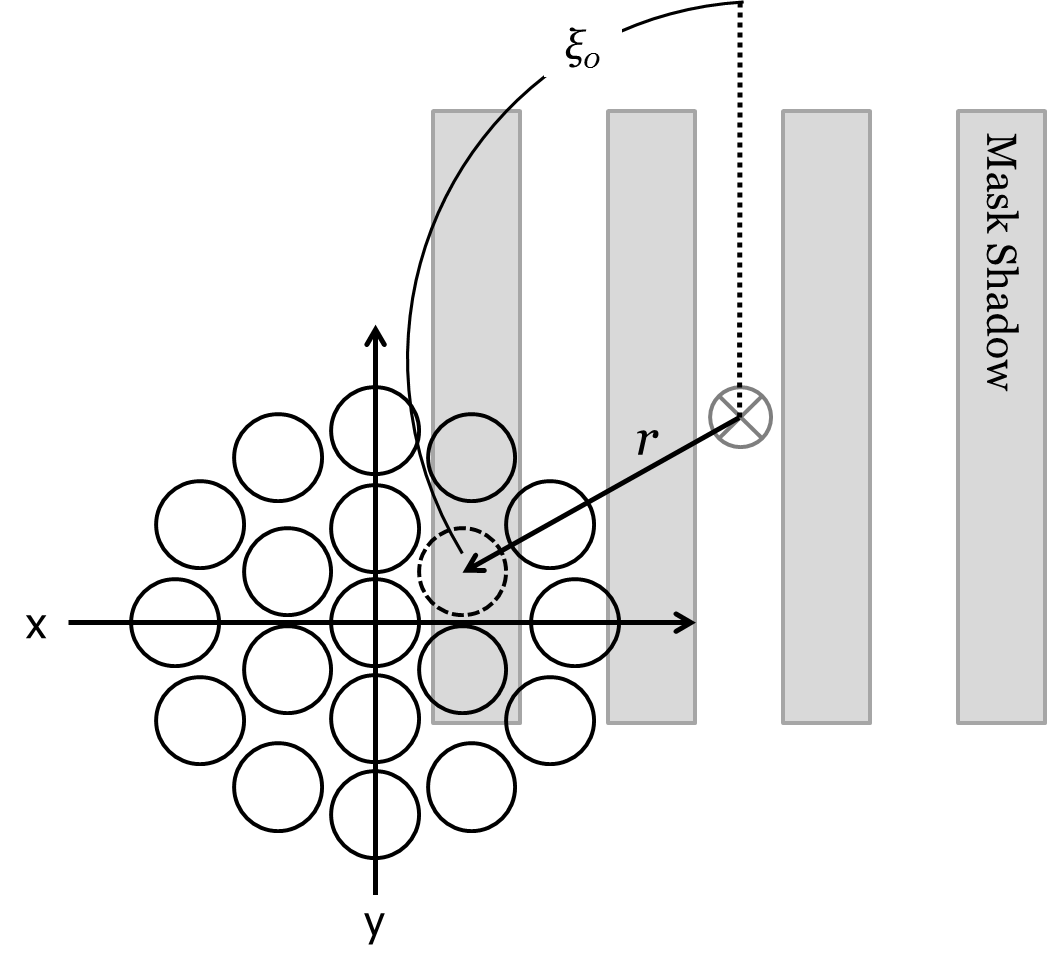} \\
	\end{center}
	\caption[example]
   { \label{fig:detloc} 
   Top view diagram of the RM, describing the polar location $(r,\xi_0)$ of a particular detector (dashed outline) relative to the mask shadow origin ($\bigotimes$).}
\end{figure}

A point source in the object scene has intensity $I_0$ and is located at azimuth $\phi$ and zenith angle $\theta$. For a detector centered at $(x_0,y_0)$ in the lab frame (relative to an origin coincident with the mask's rotational axis), the detector's polar coordinates $(r,\xi_0)$ relative to the shadow's axis projected through the mask (Fig. \ref{fig:detloc}) are given by
\begin{equation}
r = \sqrt{ (x_0 + L \tan\theta \cos\phi)^2 + (y_0 + L\tan\theta\sin\phi)^2 },
	\label{eq:r_basic}
\end{equation}
and
\begin{equation}
	\xi_0 = \tan^{-1}\left[ \frac{y_0+L \tan\theta\sin\phi}{x_0+L \tan\theta\cos\phi} \right].
	\label{eq:xi_basic}
\end{equation}

The mask's angular orientation is $\xi(t) = \omega t$ for frequency $\omega$. The x-component of the leading edge of the first slat in the $-\hat{x}$ direction from the origin is given by
\begin{equation}
	x(t) = r\cos{(\xi(t)+\xi_0)}.
	\label{eq:x_basic}
\end{equation}
To account for the periodic traversal by multiple slats, the transmission through the spacings between slats, and the symmetry of the system, a modified x coordinate is defined in units of detector diameter $a$,
\begin{equation}
	x^*(t) = 1 - \left| \left| \frac{x(t)}{a} \right| \mbox{ mod } 2 - 1 \right|,
	\label{eq:xStar_basic}
\end{equation}
The $x^*$ component defines the traversal across the detector diameter of the leading edge of whichever bar shadows the detector at time $t$. The percentage of the detector face that is shadowed is given by the integration over the area of a circle from zero to a fractional distance $\tau$ across its diameter,
\begin{equation}
	F(\tau) = \frac{1}{\pi} \cos^{-1}(1-2 \tau) - \frac{2}{\pi}(1-2\tau) \sqrt{\tau-{\tau}^2}.
	\label{eq:F_basic}
\end{equation}
The characteristic count rate profile measured by detector $d$ is found by subtracting from 100\% transmission the fractional shadowing of the detector described by Eq. \ref{eq:F_basic} (with $x^*$ given as the input variable), scaled to the intensity of the source:
\begin{equation}
	P_d(t) = I_0\left(1 - F\left[ x^*(t) \right]\right).
	\label{eq:p_basic}
\end{equation}

\section{Advanced Characteristic Count Rate Formula}
\label{sec:advanced}

\subsection{Introduction}

An advanced characteristic profile allows for greater flexibility in the mechanical design of the RM by allowing mask and detector geometry to be defined independently. It also describes attenuation, clipping, and shadow-lengthening effects as a function of photon energy, and the cumulative shadowing of a detector by multiple slats simultaneously. 

The RM slat width $a$, slat spacing $b$, and detector diameter $c$ are defined independently. Additionally, the slats have thickness $h$, while $L$ describes the distance from the detection plane to the bottom of the mask.

\subsection{Description}

\begin{figure}[!t]
	\begin{center}
		\includegraphics[width=1.8in]{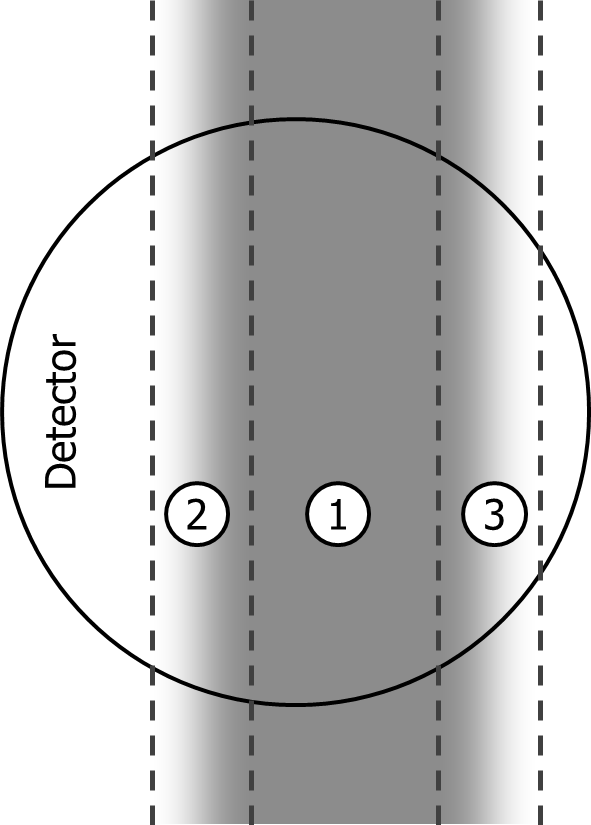} \\
	\end{center}
	\caption[example]
   { \label{fig:shadow} 
   Top-down closeup view of slat shadow on detector face. A slat shadow in the advanced formula may be divided into three regions: the ``full thickness'' region (1) where the shadow opacity is uniform, and the outer clipping sections (2,3) where the shadow opacity decreases exponentially away from region (1).}
\end{figure}

It is useful to consider the shadow from a particular slat being composed of three regions (Fig. \ref{fig:shadow}): the middle region of the shadow is due to attenuation of photons that are incident on the ``full thickness'' portion of the slat, i.e. the photon trajectory penetrates both the top and bottom face of the slat; the resulting shadow is spatially uniform. The other two regions reside on the outside of the full thickness shadow and are a result of the attenuation of photons whose trajectory ``clips'' the slats, i.e. the trajectories go through either the top or bottom face of the slats, but not both. Since the distance a photon travels through this portion of the slat varies depending on how close it is to the slat edge, the opacity of this shadow will be non-uniform with exponential decrease away from the full thickness region.

\begin{figure}[!t]
	\begin{center}
		\includegraphics[width=2.9in]{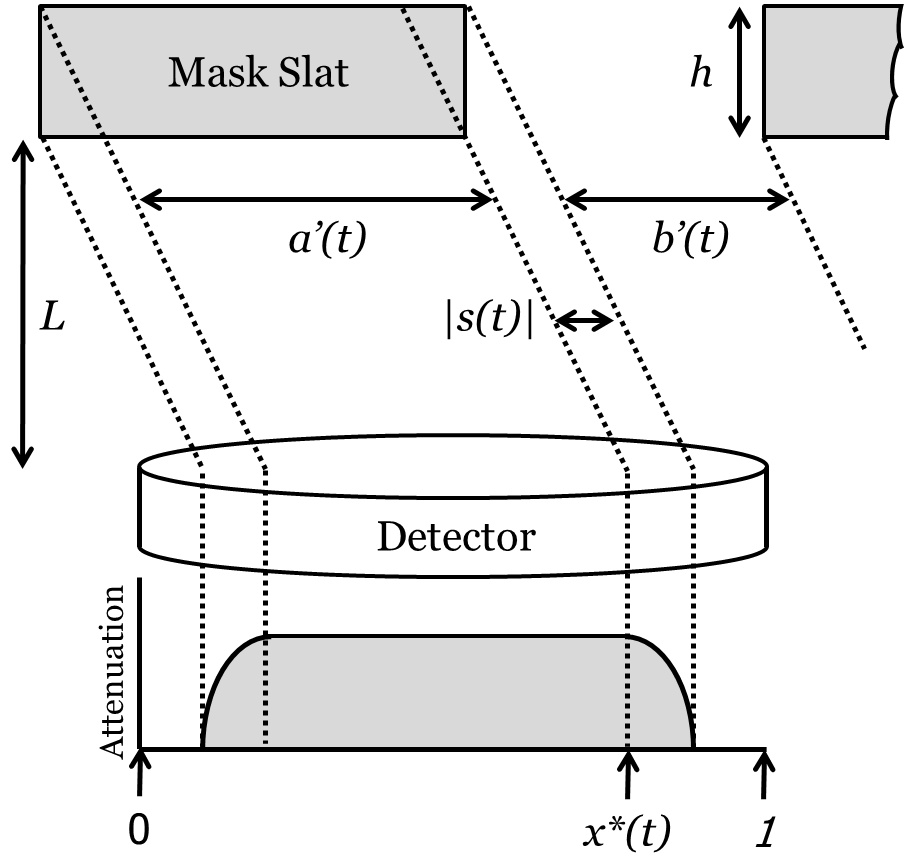} \\
	\end{center}
	\caption[example]
   { \label{fig:advanced} 
   Side view diagram of the time-variable mask shadow parameters. The projected leading (bottom) edge of the bar is given by $x^*(t)$. The full-thickness shadow region has width $a'(t)$, the clipping regions width $|s(t)|$, and $b'(t)$ describes the slat shadow spacing.}
\end{figure}

As the RM mask rotates, the width of a slat shadow will vary due to finite slat thickness by an amount $|s(t)|$, where
\begin{equation}
	s(t) = h \tan\theta \sin(\xi(t)-\phi).
	\label{eq:s}
\end{equation}
The absolute value of this parameter is the width at time $t$ of either of the two clipping shadow regions. Because of this effect, the apparent point of symmetry in the mask shadow will shift by an amount $s(t)/2$. The x-component of the leading slat shadow is modified from Eq. \ref{eq:x_basic}, to become
\begin{equation}
	x'(t) = r \cos(\xi(t) + \xi_0) + \frac{s(t)}{2}.
\end{equation}
The polar coordinates $(r,\xi_0)$ for the starting position of the detector in the frame of the projected mask shadow for a source at $(\theta,\phi)$ are still given by Eqs. \ref{eq:r_basic} and \ref{eq:xi_basic}. Similarly, the width of the ``full thickness'' shadow region will shrink relative to $a$ due to the increased percentage of incident photons that clip the slats (Fig. \ref{fig:advanced}),
\begin{equation}
	a'(t) = a - |s(t)|.
	\label{eq:aP}
\end{equation}
The coordinate $x^*$ is next defined (similar to Eq. \ref{eq:xStar_basic}) to account for the mask's periodic nature, symmetry, and in this case, the ability for multiple bars to shadow the detector face simultaneously. Consequently, $x^*$ must be an array of time-dependent functions,
\begin{multline}
	x_m^*(t) = \frac{1}{c} \left(\frac{1}{2}\left[ a'(t)+c \right] \right. \\
	\left. - \left| |x'(t)| \mbox{ mod } (a+b) - (1+2m)\frac{a+b}{2} \right| \right), \\
	-M \le m \le M.
\end{multline}
Index $m$ spans the integer values from $-M$ to $M$. The size of the array is equal to the total number of bars that may simultaneously shadow any one detector, $1+2M$, where $M$ is the integer given by\footnote{The function floor$(x)$ rounds $x$ down to the nearest integer.}
\begin{equation}
	M = \mbox{floor} \left( \frac{a+s_{max}+c}{a+b} \right).
\end{equation}
The maximum $s(t)$ value, $s_{max}$, occurs when the mask angle is $90^{\circ}$ and $270^{\circ}$ out of phase with the source azimuth:
\begin{equation}
	s_{max} = h\tan\theta.
\end{equation}

The fraction of detector area shadowed (similar to Eq. \ref{eq:F_basic}) becomes
\begin{multline}
	F_0(\tau) = \frac{1}{\pi} \cos^{-1}(1-2 \Lambda[\tau]) \\
	- \frac{2}{\pi}(1-2\Lambda[\tau]) \sqrt{\Lambda[\tau]-\Lambda[\tau]^2},
	\label{eq:F0}
\end{multline}
where we have introduced the constraint formula,\footnote{The function $\mbox{max}\{x,y\}$ returns the maximum value of the two input values, or equivalently the minimum for $\mbox{min}\{x,y\}$.}
\begin{equation}
	\Lambda[\tau] = \mbox{min}\{\mbox{max}\{\tau,0\},1\}.
\end{equation}
In the standard formula in the previous section, this constraint is not necessary since only a single slat is evaluated, and its width equals the detector diameter. Since multiple bars can now contribute to the shadowing, the constraint limits the integration of shadow between the two edges of the detector area.

Using Eq. \ref{eq:F0}, the fractional shadowing of the detector from all contributing slats is then
\begin{multline}
	G_0(t) = \left(1 - \mbox{exp}\left[-\frac{h\sigma\rho}{\cos\theta}\right]\right) \left[ F_0\left(x^*_m(t) + \frac{|s(t)|}{c}\right) \right. \\
		\left. - F_0\left( x^*_m(t) - \frac{a'(t)}{c} - \frac{|s(t)|}{c} \right)  \right],
	\label{eq:G0}
\end{multline}
where $\sigma$ and $\rho$ are the mass attenuation coefficient (for a particular photon energy) and density of the slat material, respectively. $G_0$ includes the shadow contributions from all three regions of the slat shadow with constant attenuation assumed. To account for the additional transmission through the clipping portion of the shadow, a function $F_1$ is defined,
\begin{multline}
	F_1(\tau,t) = \\
	\frac{4}{\pi} \int_{\Lambda[\tau]}^{\Lambda[\tau+|s(t)|/c]} \mbox{exp}\left[-Z(t)(\tau-x)\right] \sqrt{1 - (2x-1)^2} \mbox{ d}x \\
		+ F_0(\tau) - F_0\left( \tau + \frac{|s(t)|}{c} \right),
	\label{eq:F1}
\end{multline}
where
\begin{equation}
	Z(t) = \frac{h c\sigma\rho}{|s(t)| \cos\theta}.
	\label{eq:Z}
\end{equation}
The first part of Eq. \ref{eq:F1} is the integration of the transmission fraction of the exponentially-decreasing shadow opacity about the circular geometry of the detector; the second part removes the transmission (or lack of attenuation) already accounted for in Eq. \ref{eq:G0}. The transmission by clipped photons not accounted for in Eq. \ref{eq:G0} is given by
\begin{multline}
	G_1(t) = \mbox{exp}\left[-\frac{h\sigma\rho}{\cos\theta}\right] \bigg[ F_1(x^*_m(t),t) \bigg. \\
	\bigg. + F_1\left(1-x^*_m(t)+\frac{a'(t)}{c},t\right) \bigg].
\end{multline}

The advanced characteristic count rate profile combines the above functions:
\begin{equation}
	P_d(t) = I_0 \left( 1 - \sum_m \left[ G_0(t) - G_1(t) \right] \right).
	\label{eq:Padv}
\end{equation}

\subsection{Approximation for Practical Use}

Equation \ref{eq:F1}, describing the integration of the exponential decrease in shadow opacity across the detector, has no closed-form solution and is computationally expensive to evaluate numerically. Also, as the source azimuth $\phi$ and grid angle $\xi(t)$ align, $s(t) \rightarrow 0$ and $Z(t) \rightarrow \infty$; the span of the integral in $F_1$ thus approaches zero causing numerical solutions of $F_1$, and consequently $G_1$, to become unstable.

In most cases, the $G_1$ transmission contribution from the clipped photons will be a small contribution relative to $G_0$, and so may be simply ignored in Eq. \ref{eq:Padv}. The shadow-lengthening effects are still described by $G_0$ with uniform attenuation assumed for the clipping shadow regions. A suitable approximation for $G_1$, however, is found by using the ratio of the integral underneath the isolated exponential function in Eq. \ref{eq:F1} to the full transmission ignoring the shape of the detector,
\begin{equation}
	\alpha = \frac{\cos\theta}{h\sigma\rho} \left( 1 - \mbox{exp}\left[-\frac{h\sigma\rho}{\cos\theta}\right] \right) - \mbox{exp}\left[-\frac{h\sigma\rho}{\cos\theta}\right].
	\label{eq:alpha}
\end{equation}
A fortunate result of this approach is that $\alpha$ is constant over $s(t)$, and thus all grid angles. This provides a computationally fast solution for approximating $G_1$, which is written as
\begin{multline}
	\widetilde{G_1}(t) = \alpha \left[ F_0\left(x^*_m(t) + \frac{|s(t)|}{c} \right)  \right. \\
	- F_0\left( x^*_m(t) \right) + F_0 \left( x^*_m(t) - \frac{a'(t)}{c} \right)  \\
	\left.  - F_0 \left( x^*_m(t) - \frac{a'(t)}{c} - \frac{|s(t)|}{c} \right) \right].
\end{multline}

\subsection{Limitations of Use}

While this advanced characteristic formula for the RM is an improved representation of the instrument response over the standard formula, it does have limitations to its use. It has been designed specifically to allow for higher-energy photons and a larger FOV. These two properties effectively work against one another, however, since according to Eqs. \ref{eq:s} and \ref{eq:aP}, as the thickness $h$ or source azimuth $\theta$ increase, $a'(t)$ decreases. In order to maintain the integrity of the advanced formula, it is required that min$\{a'(t)\} \ge 0$. Otherwise, some photon trajectories may pass through both sides of a slat, but not the top or bottom faces, which the advanced formula does not describe.

Similarly, if we define a parameter $b'(t)$ as the time-dependent distance from the outermost edge of one slat shadow to the next (see Fig. \ref{fig:advanced}), 
\begin{equation}
	b'(t) = b - |s(t)|,
\end{equation}
then we must also ensure that $\mbox{min}\{b'(t)\} \ge 0$; else, illumination of the detector may never be achieved. These two requirements provide a constraint on the geometry of the mask subject to the FOV, $\theta_{FOV}$, of the instrument:
\begin{equation}
	a \ge h\tan\theta_{FOV}, \quad b \ge h\tan\theta_{FOV}.
\end{equation}

\section{Simulation \& Results} \label{sec:results}

\begin{figure}[!t]
	\begin{center}
		\includegraphics[width=3.3in]{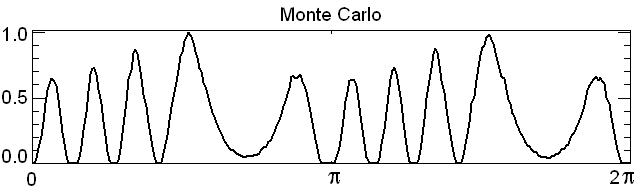} \\
		\includegraphics[width=3.3in]{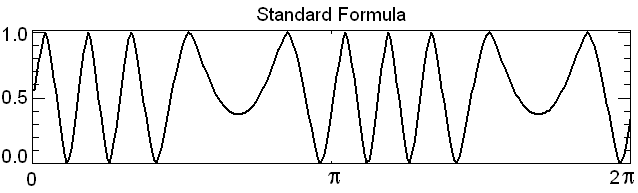} \\
		\includegraphics[width=3.3in]{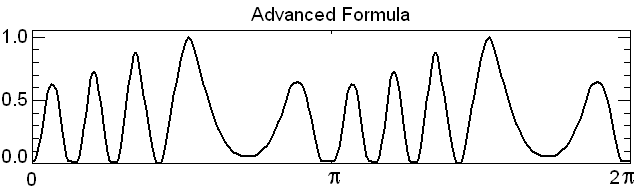}
	\end{center}
	\caption[example]
   { \label{fig:shadlength}
   Count profiles for a mask with slat thickness 20 cm, source at $\theta = 6^{\circ}$, and $a=b=c=4$ cm. The horizontal axis is given in terms of mask rotation angle $\xi$, while the vertical axis is measured source intensity in arbitrary units. The result of shadow lengthening due to the large mask thickness is evidenced by the asymmetry in the profile.}
\end{figure}

We perform Monte Carlo simulations with various instrument geometries and compare the results with the profiles derived from the advanced and, where suitable, the standard count rate formulae. For the results described below (unless otherwise indicated), a lead mask ($\rho = 11.34$ g/cm$^3$) is suspended $L = 1$ m above the detection plane. Monoenergetic 662 keV photons have a total mass attenuation coefficient\footnote{NIST XCOM, http://www.nist.gov/pml/data/xcom/} of $\sigma = 0.103$ cm$^2$/g. Results are computed for various combinations of $a$, $b$, and $c$. Background is assumed to be zero, and only photopeak events are included in the analysis. Additional mask geometry and source parameters are selected for each scenario individually to demonstrate a particular advantage of the advanced formula.

\subsection{Computational Speed}

For a direct comparison of the computational expense for the advanced versus the standard formula, instrument response functions are calculated using both solutions for an RM with a 14$^{\circ}$ FOV divided into 12$'$ field bins (4900 elements) and count profiles broken up into 560 time bins. The calculations are performed using IDL 6.3 on a Windows machine.

\begin{figure}[!t]
	\begin{center}
		\includegraphics[width=3.3in]{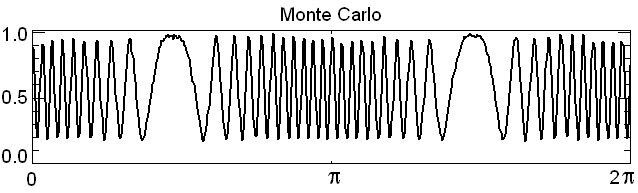} \\
		\includegraphics[width=3.3in]{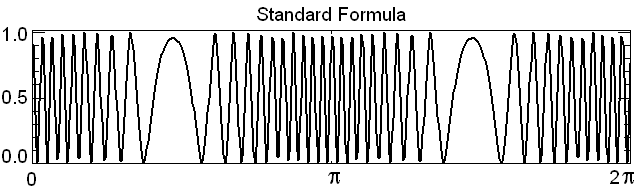} \\
		\includegraphics[width=3.3in]{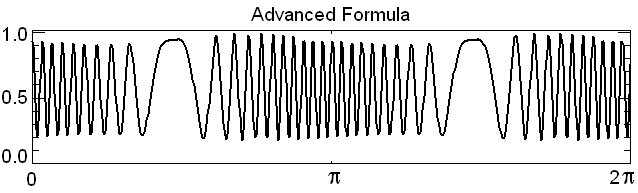}
	\end{center}
	\caption[example]
   { \label{fig:shadlength2}
   Count profiles for a mask with slat thickness 1 cm and source at $\theta = 50^{\circ}$. Shadow lengthening due to the large incident angle is seen as an asymmetry in the envelope of the high frequency modulations, and the lack of attenuation due to the thin mask appears as a reduction of contrast in the profile.}
\end{figure}

The standard formula computes the instrument response (one profile per sky bin) in 0.8 s for each detector. The advanced formula, ignoring the $G_1$ term, takes 2.6 s, while inclusion of the approximated $\widetilde{G_1}$ term increases the time to 5.4 s. The processing time is still many orders of magnitude shorter than that required to determine the instrument response using a Monte Carlo simulation. If we require a signal-to-noise ratio of 10 per time bin to derive a Monte Carlo profile that is suitable for the purposes of image reconstruction, the intrument response for a single detector would take almost 1.3 days to compute.



\subsection{Count Rate Profiles}

The count rate profiles as calculated by the two formulae are compared directly to the data recorded with Monte Carlo simulations. Two scenarios are first examined with $a = b = c = 4$ cm, so that the advanced formula can be compared directly to the standard formula (where $a=b=c$ is required). Shadow lengthening is demonstrated in Fig. \ref{fig:shadlength} by increasing the thickness of the slats to 20 cm and placing a source at zenith angle $\theta = 6^{\circ}$. The Monte Carlo result shows a profile that is asymmetric, with transmission peaks varying in height due to the broadened shadow. The standard formula is incapable of accurately describing this effect, since it assumes infinitesimal slat thickness; the advanced profile, however, is virtually identical to the Monte Carlo result. Though the thickness of the slats here is greatly exaggerated, the purpose is to demonstrate the necessity of an accurate description of the instrument response with a thick mask to image higher energy gamma rays.

Shadow lengthening is again examined in Fig. \ref{fig:shadlength2} by instead increasing the source zenith angle to 50$^{\circ}$ with a slat thickness of only 1 cm. Two features of the Monte Carlo profile are observed here: (1) the asymmetry in the envelope of high frequency modulations between the two low-frequency peaks and (2) the decreased profile contrast (i.e., a minimum count rate which does not go all the way to zero) due to transmission through the mask. The standard formula is unable to describe either of these features, and so would be unsuitable for analysis with an instrument that has a large field of view.

\begin{figure}[!t]
	\begin{center}
		\includegraphics[width=3.3in]{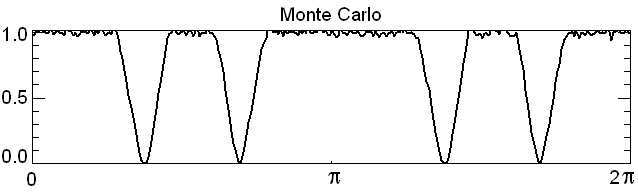} \\
		\includegraphics[width=3.3in]{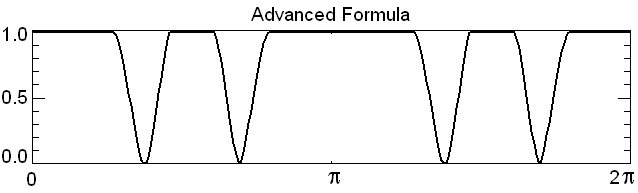}
	\end{center}
	\caption[example]
   { \label{fig:incTrans}
   Spacing between mask slats $=$ 12 cm, while slat and detector widths $=$ 4 cm. The increased transmission is well represented by the advanced formula, as the profile intensity saturates several times during the rotational period.}
\end{figure}

We next examine the results of altering the mask geometry by removing the $a=b=c$ constraint. Since the standard formula utilizes only a single variable to account for these three mask parameters, a direct comparison is not possible. Instead, only the Monte Carlo and advanced profiles are presented. The profiles for an RM with mask geometry providing increased transmission is shown in Fig. \ref{fig:incTrans}. The slat spacing is increased to $b=12$ cm, while keeping the slat and detector widths at 4 cm. The advanced profile demonstrates the expected saturation of the transmission intensity several times during the rotational period. 

\begin{figure}[!t]
	\begin{center}
		\includegraphics[width=3.3in]{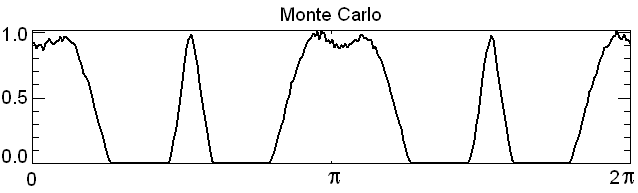} \\
		\includegraphics[width=3.3in]{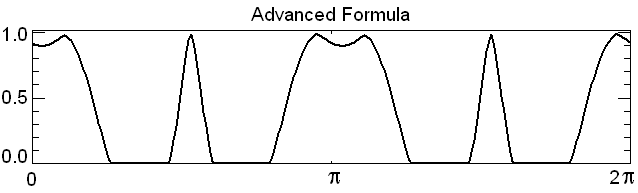}
	\end{center}
	\caption[example]
   { \label{fig:incShad}
   Slat widths $=$ 12 cm, with slat spacing and detector diameters $=$ 4 cm. The increased shadowing is seen as intervals of zero-intensity transmission. }
\end{figure}

\begin{figure}[!t]
	\begin{center}
		\includegraphics[width=3.3in]{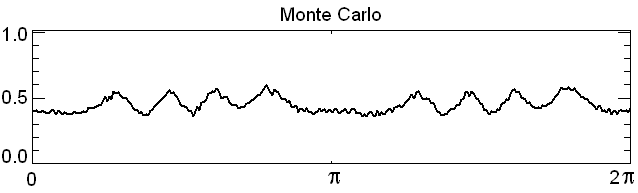} \\
		\includegraphics[width=3.3in]{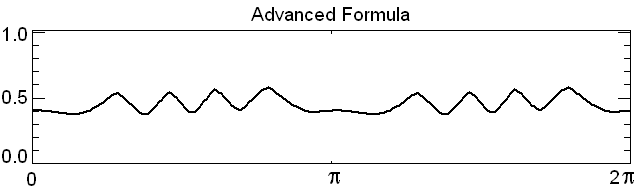}
	\end{center}
	\caption[example]
   { \label{fig:multBars}
   For slat width and spacing $=$ 4 cm and detector diameter $=$ 12 cm, up to two bars may simultaneously shadow the detector.}
\end{figure}

Conversely, increased shadowing is shown in Fig. \ref{fig:incShad} by setting the slat widths to $a=12$ cm, and keeping the slat spacing and detector diameters at 4 cm. Again, the advanced formula accurately describes the ``dead time'' of the profile, with several periods of zero intensity transmission. Finally, the ability of the advanced formula to decribe the simultaneous shadowing of multiple slats is shown in Fig. \ref{fig:multBars}. Here, the slat width and spacing is 4 cm, while the detector diameter is increased to 12 cm. While the ratio of these dimensions is again extreme for purposes of demonstration, the ability to individually define these variables is key to optimization of the instrument geometry, as described in Sec. \ref{sec:extcons}, particularly when high energy sensitivity and large FOV are desired. Examples of the performance with dimensions $a=b=c$ chosen to be those of a practical laboratory prototype RM \citep{BuddenSORMA2010} are described in the next section.

\subsection{Image Reconstruction}

The ultimate goal of solving the system of equations in Eq. \ref{eq:oMatrix} is to produce an accurate reconstruction of the object scene. We therefore now compare the reconstructed images of Monte Carlo data based on both the standard and advanced characteristic count rate formulae. We use an image reconstruction technique specifically developed for the RM that is capable of achieving ``super-resolution'' (resolving power better than the ratio of slat spacing to mask-detector separation), while compensating for statistical noise. The algorithm uses an iterative algebraic solution with physical constraints as a form of non-linear regularization and the addition of appropriate randomized noise layers to suppress spurious background. Details of the Noise Compensating Algebraic Reconstruction (NCAR) algorithm are given in \cite{BuddenTNS2010}. In \cite{BuddenSORMA2010}, it is shown with laboratory measurements that an RM featuring an array of 19 3.8 cm diameter $\times$ 2.5 cm thick scintillating detectors (approximately the dimensions $a=b=c=4$ cm used in the simulations of Figs. \ref{fig:shadlength} and \ref{fig:shadlength2}) is capable of resolving two point sources separated by 35$'$ in the presence of background, where the geometric resolution ($\Delta\theta = b/L$) of the instrument is 1.9$^{\circ}$.

\begin{figure}[!t]
	\begin{center}	\includegraphics[width=1.09in]{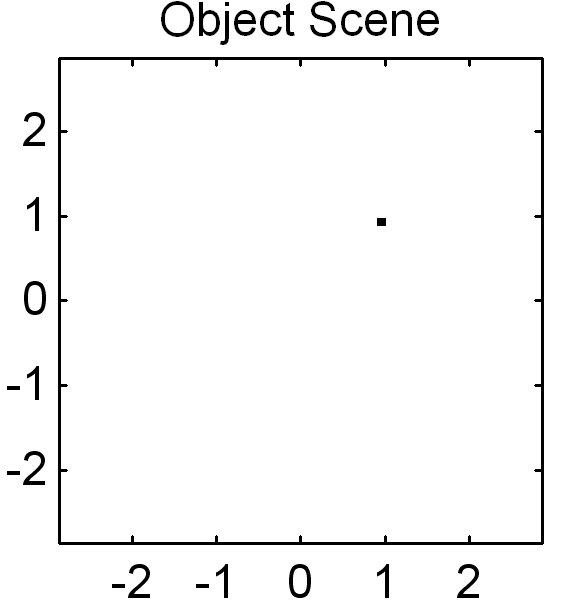}\includegraphics[width=1.09in]{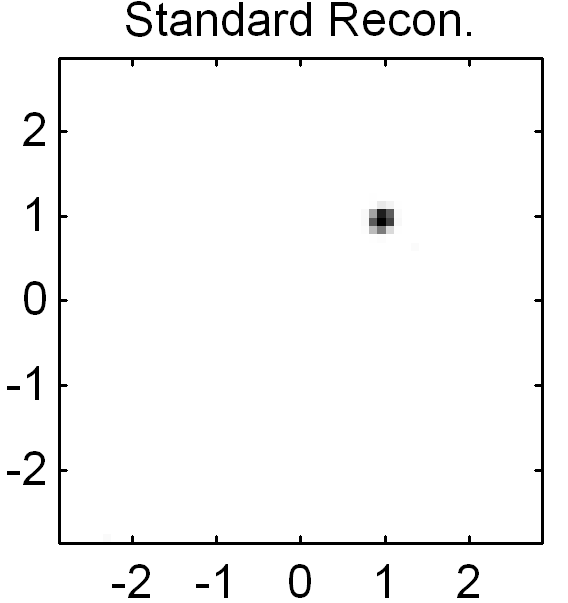}\includegraphics[width=1.09in]{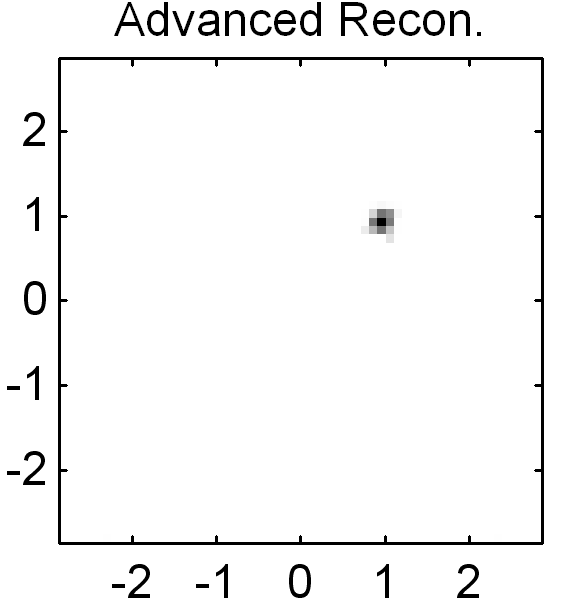}
	\end{center}
	\caption[example]
   { \label{fig:image_basic}
   A 122 keV point source is imaged by an RM with mask thickness 2 cm. The standard and advanced formula reconstructions both accurately depict the object scene. (Axes in degrees.)}
\end{figure}

First, if a 122 keV point source at small zenith angle is imaged by an RM with mask thickness 2 cm, providing 100\% attenuation, both the standard and advanced formulae produce accurate image reconstructions (Fig. \ref{fig:image_basic}). The slightly extended reconstructions reflect the locational uncertainty of the source due to noise in the data, as explained in \cite{BuddenTNS2010}.

\begin{figure}[!t]
	\begin{center}	\includegraphics[width=1.09in]{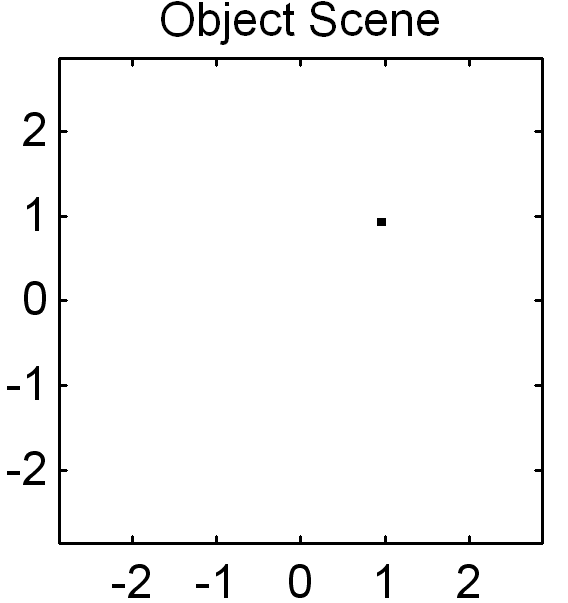}\includegraphics[width=1.09in]{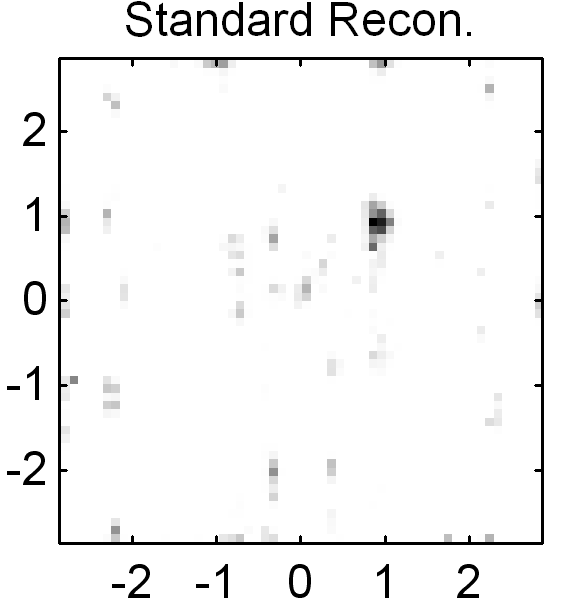}\includegraphics[width=1.09in]{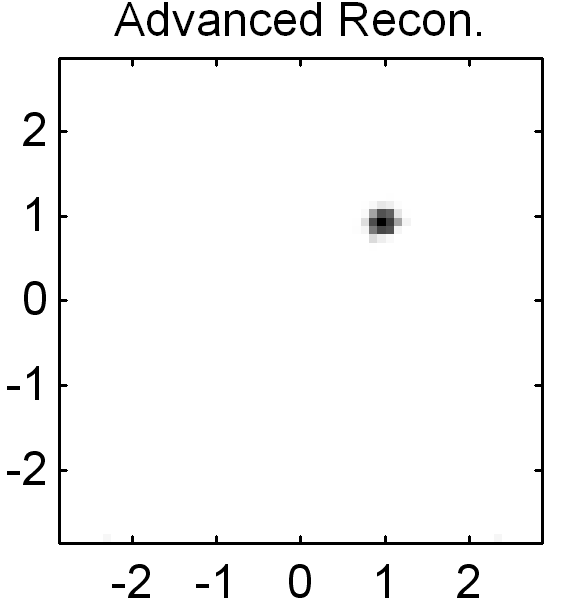}
	\end{center}
	\caption[example]
   { \label{fig:image_thin}
   A 5 mm-thick mask is used to reconstruct a 662 keV point source. Due to increased transmission through the mask, the standard formula reconstructs additional noise in the object field that is not present in the advanced formula reconstruction. (Axes in degrees.)}
\end{figure}

Next, a 5 mm mask is used to image a 662 keV point source in the same location, allowing 55\% transmission of photons incident on the slats (Fig. \ref{fig:image_thin}). The reduced contrast is properly accounted for in the advanced formula. In the image based on the standard formula, however, the increased transmission is reconstructed as spurious peaks in the image.

\begin{figure}[!t]
	\begin{center}	\includegraphics[width=1.09in]{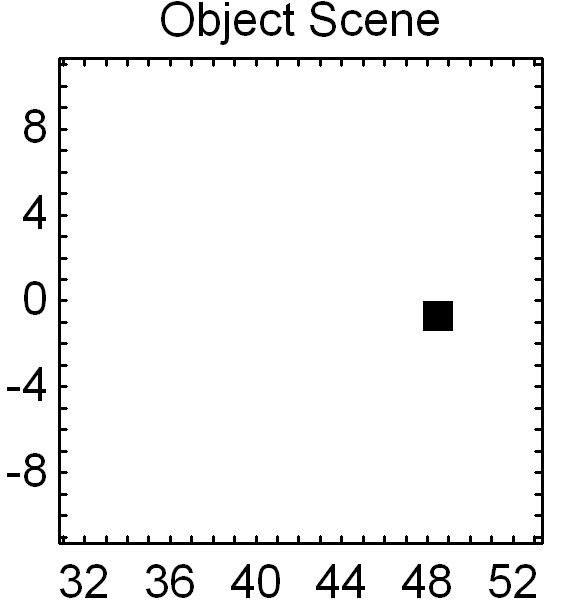}\includegraphics[width=1.09in]{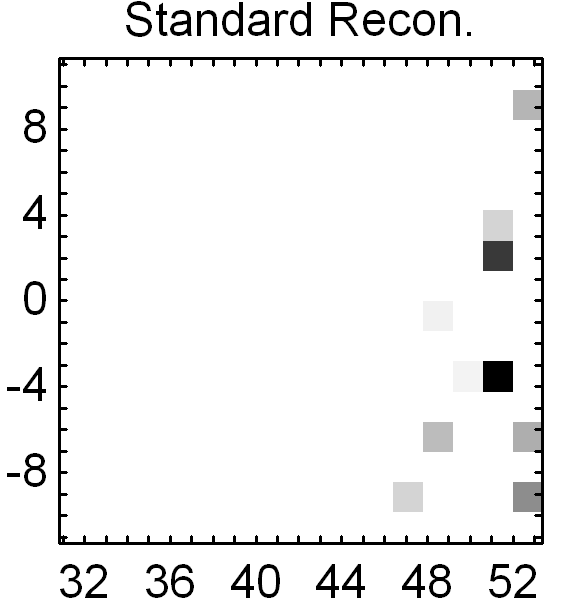}\includegraphics[width=1.09in]{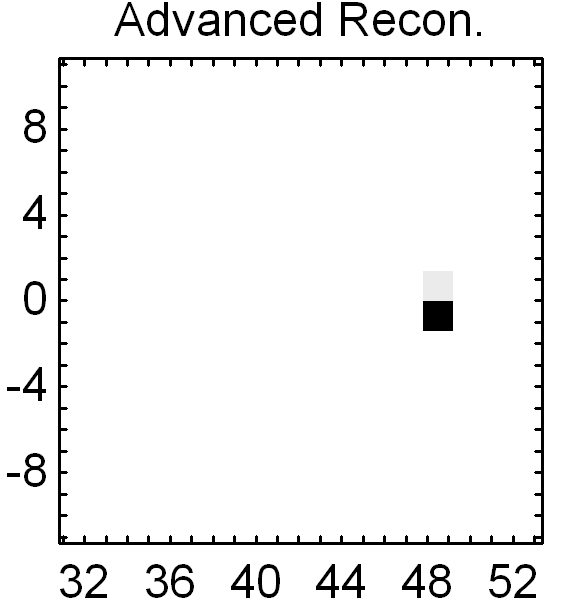}
	\end{center}
	\caption[example]
   { \label{fig:image_largezen}
   A subsection of a full reconstruction for an RM with mask 2 cm, and a point source located at $\theta \approx 48^{\circ}$. At this large zenith angle, the standard formula is incapable of correctly resolving the source. Multiple spurious peaks are visible. (Axes in degrees.)}
\end{figure}

Finally, the 662 keV source is moved to a large zenith angle, $\theta \approx 48^{\circ}$, and imaged by an RM with a 2 cm thick mask (Fig. \ref{fig:image_largezen}). At such a large angle of incidence, a large portion of photons will clip the slats, and so this effect must be accurately represented in the instrument response formula. Since the standard formula does not take this into account, the reconstruction contains spurious peaks, and no source in the true location. The advanced formula, however, reconstructs the point source in the expected location and virtually free of noise.

The effect of shadow lengthening is poorly described by the standard formula, and so the respective reconstruction is misrepresented and mislocated. The advanced formula, however, provides an accurate reconstruction of the source with high fidelity.

\section{Discussion}
\label{sec:extcons}

We have demonstrated that the advanced formula performs better than the standard instrument response by more accurately reconstructing the object scene. For a typical RM, where $a=b=c$, this advantage is most apparent for high photon energies (requiring a thick mask) and for large angles of incidence. In these situations, slat clipping and shadow lengthening have a significant effect on the profile shape, which must be accounted for in image reconstruction.

In the standard formula, the shadow opacity is uniform and its width assumed to remain unchanged despite the source zenith. For this reason, the optimal values for $a$ and $b$ equal $c$. The equality ensures two conditions are met: (1) the profile has maximum contrast (varies from 0 to 100\%), and (2) there is zero ``dead time'' (time intervals with no modulation). The former condition specifies $a \ge c$ and $b \ge c$ to maximize the standard deviation of the profile, which we have previously shown \citep{BuddenTNS2010} to be directly proportional to the sensitivity of the instrument. Given these inequalities, the latter condition then maximizes transmission with $a=c$ and resolving power with $b=c$.

The advanced formula substitutes static values $a$ and $b$ with temporally-dependent parameters, $a'(t)$ and $b'(t)$. Consequently, there is no value for $a$ and $b$ which simultaneously maximizes transmission, resolving power, and standard deviation as with the standard formula. Rather, a tradeoff must be made based on the goal and application of imaging. The advanced formula may thus be used to obtain the geometrical parameters (namely slat width and spacing) for the desired optimization.

\section{Conclusion}

An RM is an instrument capable of imaging photons in the hard x-ray and soft gamma-ray spectrum. As a mask of opaque slats rotates above a small array of non-imaging detectors, the observed count rate from the object scene is temporally-modulated, and so a time-history of counts is recorded by each detector. Subsequent folding and processing of the data can then reconstruct the object scene. To perform the deconvolution, however, the instrument response must be pre-determined and well-known. The instrument response, which is a collection of count rate profiles for all possible point sources in the object scene, is determined analytically in order to minimize computation time. 

The standard characteristic count rate formula constrains the mask and detector geometry, and is incapable of describing complex attenuation effects which are important at relatively high energies. A more robust characteristic formula to describe the instrument response is essential to the design of a wide-field high resolution RM telescope suitable for high-energy x-ray or gamma-ray astronomy from a long-duration balloon or satellite payload. We have presented an advanced characteristic formula that provides the expected instrument response for a flexible mask geometry and is capable of describing non-uniform attenuation, clipping effects, shadow lengthening during the exposure, and the simultaneous shadowing of a detector by multiple slats.

We have demonstrated the improved accuracy of this advanced formula by comparison to the standard formula and Monte Carlo simulation results. The profiles determined with the advanced formula are a visibly better representation of the Monte Carlo results, and the reconstructed images a more accurate depiction of the object scene. The profiles also allow for the derivation of optimal mask geometry to maximize sensitivity, resolving power, or transmission, based on desired imaging characteristics and application.

%
\acknowledgments
This work supported in part by NASA/Louisiana Board of Regents Cooperative Agreement NNX07AT62A. B. Budden wishes to thank the Louisiana Board of Regents under agreement NASA/LEQSF(2005-2010) - LaSPACE, NASA/ LaSPACE under grant NNG05GH22H, and the Coates Foundation at Louisiana State University for support during this project.


%

%

\end{document}